\documentclass[12pt,twoside]{article}
\usepackage{fleqn,espcrc1,epsfig,wrapfig}

\usepackage{mathptmx}
\begin {document}

\title{Deuteron Compton scattering and
 electromagnetic polarizabilities of the nucleon\thanks
{Supported in part by the Advance Research Foundation of Belarus and
 by the Russian Foundation for Basic Research (grant 98-02-16534).}}

\author{
M.I. Levchuk\address
{Stepanov Institute of Physics, Scaryna prospect 70, Minsk 220072, Belarus}
and
A.I. L'vov\address
{Lebedev Physical Institute, Leninsky prospect 53, Moscow 117924, Russia}}

\maketitle

\bigskip

\begin{abstract}

Differential cross section of deuteron Compton scattering has been
calculated using the Bonn NN-potential, a consistent set of
meson-exchange currents and seagulls, and lowest- and higher-order
electromagnetic polarizabilities of the nucleon. Estimates of the
polarizabilities of the neutron are obtained from recent experimental
data.

\end{abstract}

\bigskip

Photon scattering is a powerful tool for probing the structure of
hadrons and nuclei. In particular, the electric and magnetic
polarizabilities of the proton, $\alpha_p$ and $\beta_p$, have been
successfully determined from data on low-energy $\gamma p$ scattering
\cite{gold60,bara74,macg95}.  An extension of this method to the case
of the neutron is of considerable interest.  One of suggested ideas was
to use (elastic) Compton scattering off deuterons.  The latter process
is not easy for measurements and interpretations.  On the experimental
side, one has to carefully separate elastic and inelastic channels.  On
the theoretical side, one has to separate effects caused by the
proton's and neutron's structure from effects caused by the nuclear
environment. Recently, a progress was made in solving both the
problems. First data on low-energy $\gamma d$ scattering came from
Urbana and Saskatoon \cite{luca94,horn00} and several groups reported
their calculations \cite{wilb95,levc95,kara99,levc00,chen98,bean99}.

Two different approaches were applied in the quoted theoretical works.
In \cite{wilb95,levc95,kara99,levc00}, a phenomenological picture of a
meson-mediated NN potential was used. It was assumed that the nucleons
themselves are not modified. All binding effects were described by
meson-exchange currents associated with the NN potential (and with
excitation of the $\Delta$ isobar too) and by intermediate rescattering
of nucleons between two acts of electromagnetic interaction.

In \cite{chen98,bean99}, a technique of effective field theories (EFT)
was applied.  In this approach, the very pion cloud which mediates the
NN interaction affects properties of the nucleon (like the magnetic
moment and polarizabilities), which therefore become dependent on the
nuclear environment. An attractive feature of the EFT method is that
the retardation effects, which are not inherent to the potential
picture and which are especially large in the case of the pion
exchange, are naturally included into the calculation. A big problem of
EFT is, however, that an expansion over a small generic momentum $Q$
utilized there is not rapidly convergent. In order to extract
polarizabilities of nucleons from $\gamma d$ scattering, it is not
sufficient to calculate only the leading-order and next-to-leading
order terms. Higher orders are not negligible at all (see a discussion
in \cite{levc00}). If so, numerious unknown parameters (low energy
constants) of short-range interactions which always appear in higher
orders make the strictness of the EFT illusive.  In this situation the
potential approach remains, in our opinion, more precise.  Though
model-dependent, it does not avoid facing the short-range dynamics,
including resonance exitation, when relevant.

In our calculation \cite{levc00}, the nonrelativistic version of the
Bonn one-boson-exchange-potential was used. Meson-exchange currents and
seagulls were found via photon coupling to intermediate mesons and
meson-baryon verticies. A special attention was paid to implementing
effects of the meson-baryon form factors in a gauge-invariant way.
Additional contributions due to $\Delta$-isobar excitation and
retardation effects in the pion propagator were also taken into
account. The long-range pion exchange was found to almost saturate the
meson-exchange currents and seagulls as seen in Compton scattering,
whereas short-range contributions given by heavy mesons ($\rho$,
$\omega$, $a_0$, $\sigma$) and form factors almost cancel each other.

Photon interaction with free nucleons was described keeping
leading-order relativistic corrections including the spin-orbit term
and an additional spin-independent seagull of order $1/M^3$.  Apart
from the dipole polarizabilities $\alpha_N$ and $\beta_N$ of the
nucleon, higher-order polarizabilities (spin, quadrupole, and
dispersion ones) have been taken into account as well. Their effect was
estimated using predictions of dispersion relations
\cite{babu98,lvov99}.  It was found numerically important and giving a
large positive shift in the extracted electric polarizability of the
neutron of about $+5\times 10^{-4}~\rm fm^3$.

An accuracy of the theoretical calculation was checked through a
comparison with the dispersion relation at forward angle. An agreement
better than 3\% (in the amplitude) was found at energies between 20 and
100 MeV.

\begin{figure}[h]
\vspace*{-1ex}
\begin{center}
\epsfig{file=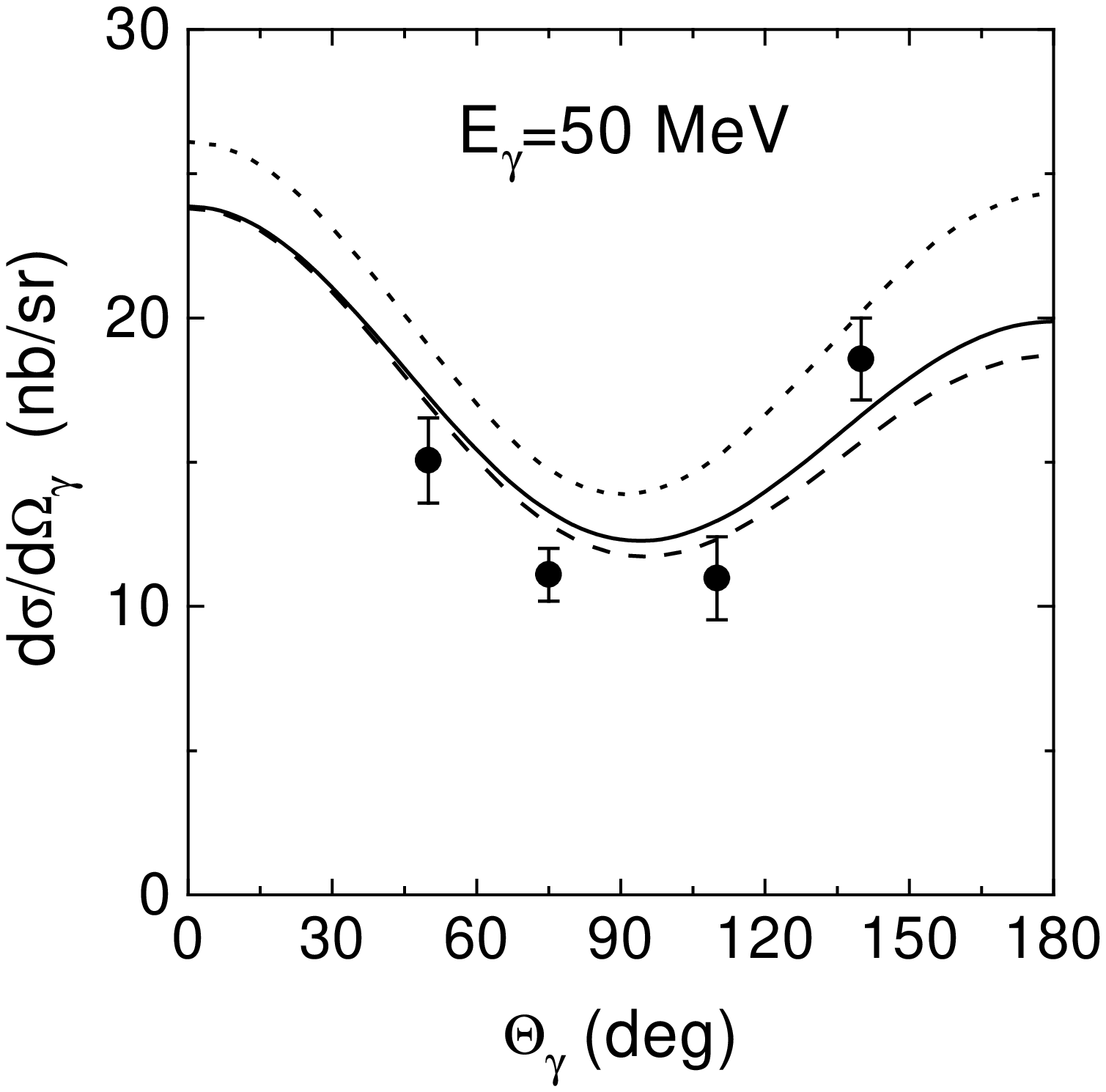,width=0.40\textwidth}
\hspace*{1cm}
\epsfig{file=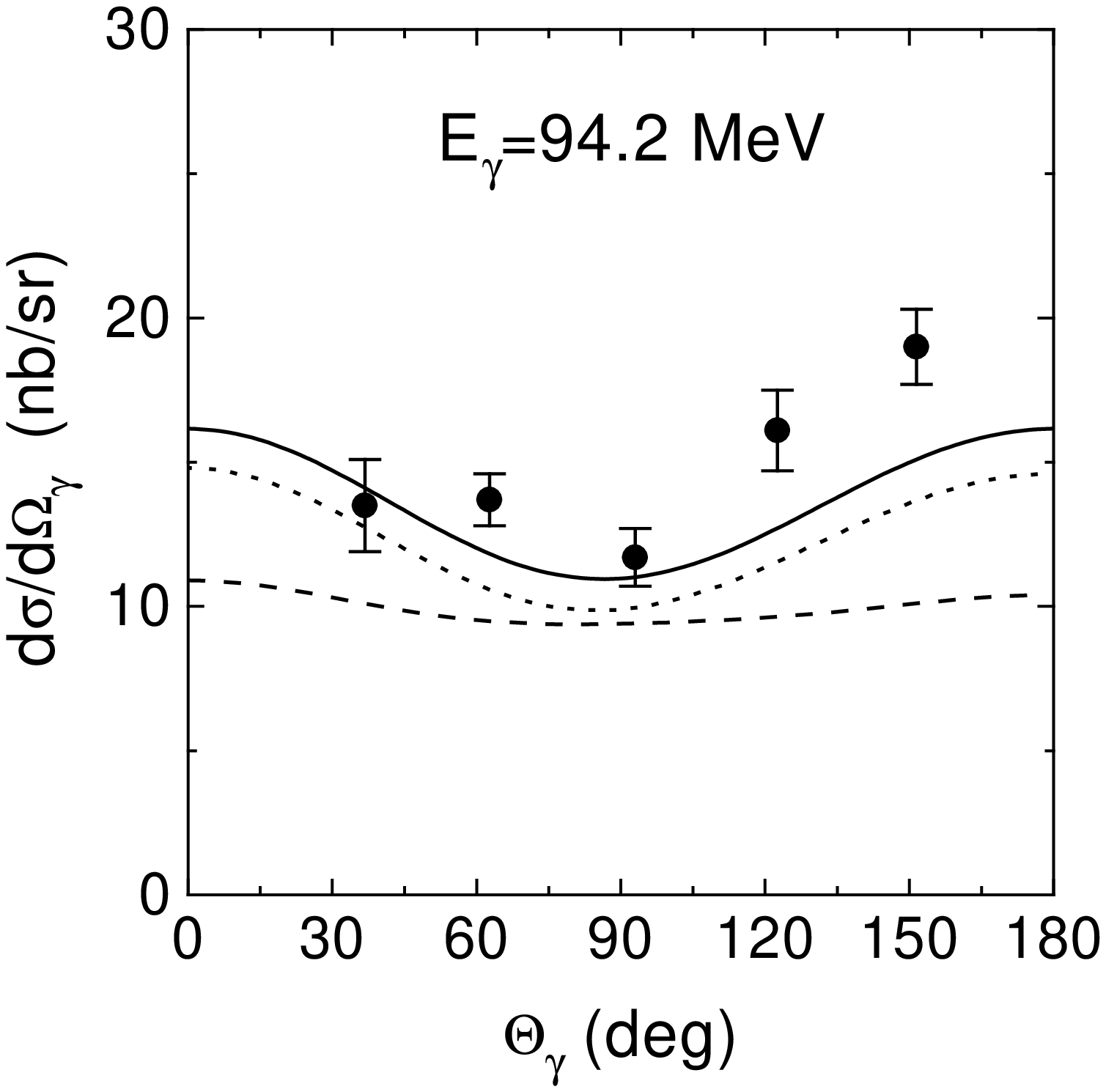,width=0.40\textwidth}
\vspace*{-7ex}
\end{center}
\caption{Differential cross section of $\gamma d$ scattering (CM).
Solid lines: present results with $\alpha_N - \beta_N = 9\times
10^{-4}~\rm fm^3$. Dashed lines: Karakowski and Miller \cite{kara99}.
Dotted lines: Beane et al.\ \cite{bean99}.  Data are from Urbana
\cite{luca94} (49 MeV) and Saskatoon \cite{horn00} (94 MeV).}
\end{figure}

Compared with other recent calculations, our results at $\sim 100$ MeV 
disagree with those obtained in \cite{kara99} (see Fig.~1).  In part, 
this disagreement is caused by a different contribution we found from 
spin-orbit terms. We have a better agreement with the EFT predictions 
from \cite{bean99}, which however becomes worser at lower energies 
$\sim 50$ MeV.  We agree at 50 MeV with \cite{kara99} and stay just 
between predictions from \cite{bean99} and \cite{chen98}.

The calculated differential cross section depends on the isoscalar
(i.e., isospin-averaged) nucleon polarizabilities $\alpha_N$ and
$\beta_N$ (see Fig.~2). Keeping all other parameters fixed, these
polarizabilities can be extracted from comparison with the experimental
data \cite{luca94,horn00}. The result is
\begin{equation}
   \alpha_N + \beta_N \approx 17.1 \pm 1.6, \quad
   \alpha_N - \beta_N \approx  4.0 \pm 1.5 \quad
   \mbox{(in units of $10^{-4}~\rm fm^3$)},
\end{equation}
where (unknown) theoretical-model errors are not yet included. Jointly
with experimental results for the proton's electromagnetic
polarizabilities, $\alpha_p \approx 12 \pm 1$ and $\beta_p \approx 2
\pm 1$ \cite{macg95}, this implies the following polarizabilities of
the neutron:
\begin{equation}
   \alpha_n \approx  9 \pm 3, \quad
    \beta_n \approx 11 \pm 3.
\end{equation}
While the obtained electric polarizability of the neutron,$\alpha_n$,
resonably agrees with the proton's polarizability $\alpha_p$ and with
dispersion-theoretical expectations (see, e.g., \cite{lvov99}), the
magnetic polarizability $\beta_n$ strongly deviates from $\beta_p$ and
from predictions of the dispersion theory which gives $\beta_n \approx
\beta_p$.  Clearly, a further work must be done before any firm
conclusion from deuteron Compton scattering can be derived concerning
the neutron's polarizabilities. In particular, of interest are
evaluations of {\em all} $1/M^2$ relativistic corrections, as well as
calculations performed with modern high-quality NN potentials.  More
precise measurements of the differential cross section of $\gamma d$
scattering at energies below pion photoproduction threshold would also
be desirable.

\columnsep=9mm
\begin{wrapfigure}{r}{7cm}
\vspace*{-3ex}
\epsfig{file=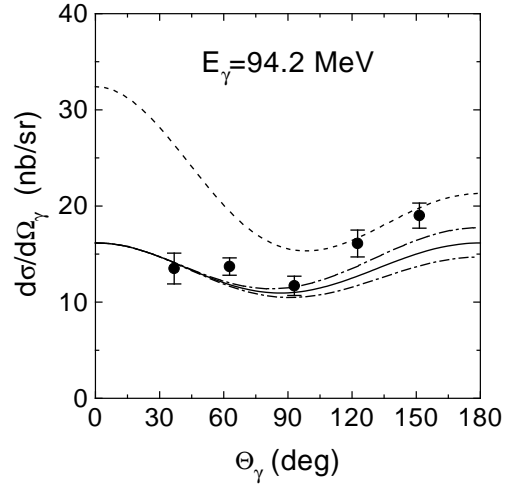,width=6.5cm}
\vspace*{-1ex}
\caption{Differential cross section of $\gamma d$ scattering (CM) at
different isoscalar polarizabilities $\alpha_N$ and $\beta_N$. Solid
line:  $\alpha_N + \beta_N = 14.6$ (inferred from the Baldin sum rule)
and $\alpha_N - \beta_N = 9$.  Units are $10^{-4}~\rm fm^3$.
Dashed-dotted lines show variations when $\alpha_N - \beta_N$ is
changed by $\pm 3$.  Dotted line: $\alpha_N = \beta_N = 0$.  Data are
from Saskatoon \cite{horn00}.}
\end{wrapfigure}

\end{document}